\documentclass[11pt,a4paper]{article}
\pdfoutput=1
\usepackage{graphicx}
\usepackage{jheppub}

\usepackage{multirow, graphicx,amssymb,url,mathrsfs,amsmath}
\usepackage{wrapfig,boxedminipage,subfigure,epsfig}
\usepackage{amsxtra,amstext,latexsym,dsfont,amsfonts}
\usepackage{color}
\usepackage[dvipsnames]{xcolor}
\usepackage{float}
\usepackage{slashed}
\usepackage{calligra}
\DeclareFontShape{T1}{calligra}{m}{n}{<->s*[2.2]callig15}{}
\DeclareMathAlphabet{\mathcalligra}{T1}{calligra}{m}{n}
\usepackage{tikz} 
\usepackage[compat=1.1.0]{tikz-feynman}
\usepackage{feynmf}
\usepackage{tensor}

\newcommand{\be}{\begin{equation}}
\newcommand{\ee}{\end{equation}}
\newcommand{\bea}{\begin{eqnarray}}
\newcommand{\eea}{\end{eqnarray}}

\newcommand{\ket}[1]{|#1\rangle}

\title{Quantum-Gravitational Null Raychaudhuri Equation}

\author[a]{Sang-Eon Bak,}
\author[a, b]{Maulik Parikh,}
\author[c]{Sudipta Sarkar,}
\author[a]{Francesco Setti}

\affiliation[a]{Department of Physics, Arizona State University, Tempe, AZ 85287, United States}
\affiliation[b]{Beyond: Center for Fundamental Concepts in Science,
Arizona State University, Tempe, Arizona 85287, USA}
\affiliation[c]{Indian Institute of Technology, Gandhinagar, Gujarat 382355, India}

\emailAdd{sbak2@asu.edu}
\emailAdd{maulik.parikh@asu.edu}
\emailAdd{sudiptas@iitgn.ac.in}
\emailAdd{fsetti@asu.edu}

\abstract{We consider a congruence of null geodesics in the presence of a quantized spacetime metric. The coupling to a quantum metric induces fluctuations in the congruence; we calculate the change in the area of a pencil of geodesics induced by such fluctuations. For the gravitational field in its vacuum state, we find that quantum gravity contributes a correction to the null Raychaudhuri equation which is of the same sign as the classical terms. We thus derive a quantum-gravitational focusing theorem valid for linearized quantum gravity.}

\begin{document}
\maketitle

\section{Introduction}
The trajectories of freely-falling test particles — geodesics — are of special importance in general relativity; among the fundamental equations in the subject are the geodesic equation for a single particle, the geodesic deviation equation for a pair of particles, and the Raychaudhuri equation for a congruence of particles. These equations pre-suppose that spacetime has a definite geometry: the spacetime metric is taken to be a classical field. A question of great interest then is: what happens to these equations when the metric is treated as a quantum field? Precisely this question was addressed in \cite{Parikh:2020nrd,Parikh:2020fhy, Parikh:2020kfh}, in which the effective dynamics of a pair of non-relativistic particles coupled to a quantized metric was obtained. After integrating out the quantized metric, one finds that the geodesic separation no longer obeys the geodesic deviation equation. Instead, even classical test particles behave stochastically, obeying a Langevin-like equation. Intuitively, the stochastic behavior of the particles arises from the probabilistic nature of the quantum field they are coupled to. This result can be obtained in different ways, either by using the Feynman-Vernon influence functional approach \cite{Feynman:1963fq} to integrate out gravity \cite{Parikh:2020nrd,Parikh:2020fhy, Parikh:2020kfh, Cho_2022}, or by treating the geodesic deviation equation itself as an operator equation \cite{Kanno:2020usf}. Subsequently, these results were extended to the case of a single non-relativistic particle falling in a weakly curved spacetime — yielding a stochastic quantum-gravity correction to the geodesic equation in its Newtonian limit \cite{Chawla:2021lop} — as well as to a congruence of massive particles, yielding stochastic quantum-gravitational corrections to the timelike Raychaudhuri \cite{Bak:2022oyn} and tidal evolution equation \cite{Cho:2023dmh}.

Here we consider the especially important case of a congruence of massless particles. When such test particles propagate in a classical spacetime, the area of the congruence evolves via the null Raychaudhuri equation, an equation of great significance in general relativity. 
Our goal in this paper is to determine how  perturbative quantum fluctuations of the spacetime geometry modify that equation in linearized quantum gravity. Previous approaches to quantize the (time-like) Raychaudhuri equation have considered the congruence as evolving non-classically because its constituent matter is quantized \cite{Ahmadi:2006nq, Das:2013oda, Vagenas:2017fwa, Choudhury:2021huy}. By contrast, the thing that is quantum in this paper is the spacetime metric; our quantum null Raychaudhuri equation therefore describes the spread of a beam of light in quantum spacetime.

Our results are as follows. We first obtain a quantum-gravity correction to the geodesic deviation equation for a pair of null particles, (\ref{full eom quantum}). As in the massive case, the coupling to a quantized metric induces random fluctuations -- noise -- to the geodesic deviation, turning the deterministic classical geodesic equation into a Langevin-like stochastic equation. The statistical properties of the noise depend on the quantum state of the gravitational field. In particular, for the Poincar\'e-invariant vacuum state, the noise contributes a term to the null Raychaudhuri equation, (\ref{eq_Quantum Raychaudhuri}), that has the same sign as the usual classical terms. This results in a quantum-gravitational focusing theorem, (\ref{quantum focusing thm}), in the regime of linearized perturbative quantum gravity.

\section{A Pair of Null Particles in a Quantized Gravitational Field} \label{sec-gravity}
Consider a null geodesic congruence in an arbitrary spacetime. We would like to determine the evolution of the area of the congruence when the spacetime metric is treated as a quantum field in some quantum field state $|\Psi \rangle$. Our strategy is to first determine how the geodesic deviation between a pair of null geodesics is affected by the quantum nature of the metric, and then to use that result to determine the behavior of a congruence of null geodesics. 
\subsection{Classical equations of motion}
The dynamics of the geodesic deviation of a pair of non-relativistic massive particles was calculated in \cite{Parikh:2020nrd,Parikh:2020fhy,Parikh:2020kfh,Cho_2022,Kanno:2020usf} and used to calculate the quantum-corrected evolution of a congruence of timelike geodesics \cite{Bak:2022oyn,Cho:2023dmh}. However, the null case is sufficiently different that it is worth starting anew from first principles. To that end, we begin by writing down the action for Einstein gravity coupled to a pair of null particles:
\begin{align}
  S = \frac{1}{ \kappa^2} \int R \sqrt{-g} d^4x +  \int d \lambda\left[\frac{1}{2 e} g_{\mu \nu} \frac{d X^\mu}{d \lambda} \frac{d X^\nu}{d \lambda}\right] + \int d \lambda' \left[\frac{1}{2 e'} g_{\mu \nu} \frac{d Y^\mu}{d \lambda'} \frac{d Y^\nu}{d \lambda'}\right]
\end{align}
where $\kappa^2=16 \pi G_N$, and $e$ and $e'$ are einbeins. Since we are interested only in the relative motion (geodesic deviation), we can regard $X$ as a reference geodesic and put $X$ and $e$ on shell. The einbein equation gives
\begin{align}
-\frac{1}{2 e^2} g_{\mu \nu} \frac{d X^\mu}{d \lambda} \frac{d X^\nu}{d \lambda}=0\,.
\end{align}
This is a constraint equation. Setting $\frac{d X^\nu}{d \lambda}$ to be null, the einbein $e$ can be arbitrary. We will consider an affine parametrization so that $\frac{d e}{d \lambda}=0$. The equation of motion of $X$ is then just the geodesic equation with affine parameterization:
\begin{align}\label{eq_geodesic eq}
\frac{d^2 X^\rho}{d \lambda^2}+\Gamma_{\mu \nu}^\rho \frac{d X^\mu}{d \lambda} \frac{d X^\nu}{d \lambda}=0
\end{align}
Since we have in mind that our two null geodesics are part of a congruence of geodesics, we can use the same parameterization for the second geodesic: $\lambda' \equiv \lambda$. 

For a sufficiently small pencil of geodesics, we can always -- by virtue of local flatness -- write the metric locally as a perturbation of the Minkowski metric:
\begin{align}
    g_{\mu\nu}(x)=\eta_{\mu\nu}+ h_{\mu\nu}(x)
\end{align}
More precisely, we can set up null Fermi normal coordinates anchored to our reference geodesic $X$. Taylor expansion of the metric in directions transverse to the null geodesic gives \cite{Blau:2006ar} 
\begin{align}\label{eq_FNC1}
    \begin{aligned}
&g_{uv}=g_{vu}=1+\mathcal{O}(x^3)\,,\\
&g_{uu}=-R_{uaub}(u)x^a x^b+\mathcal{O}(x^3)\,,\\
&g_{ua}=-\frac{2}{3}R_{ubac}(u)x^b x^c+\mathcal{O}(x^3)\,,\\
&g_{ab}=\delta_{ab}-\frac{1}{3}R_{abcd}(u)x^c x^d +\mathcal{O}(x^3)\,,
\end{aligned}
\end{align}
where we consider the light cone coordinates $u=t+z,\, v=t-z$ and the transverse coordinates  $x^a$ for $a=1,2$ with the parameter of the reference geodesic $\lambda=u$. 
Define
\begin{align}
    Y^\mu=X^\mu+\xi^\mu\,.
\end{align}
Concretely, we have $X^\mu = (u,0,0,0)$ and $Y^\mu = (u,\xi^v,\xi^a)$.
Then, to quadratic order in $\xi$, we have
\begin{align}
g_{\mu \nu} \frac{d Y^\mu}{d \lambda} \frac{d Y^\nu}{d \lambda} = -R_{uaub} \xi^a \xi^b + \frac{d\xi^v}{d u} + \delta_{ab} \frac{d\xi^a}{du} \frac{d\xi^b}{du}\,.
\end{align}
Since $\lambda = u$ and $e'(\lambda)$ is a constant (by virtue of $\lambda$ being affine), the linear term is a total derivative, and the action for the second particle thus reads
\begin{align}
\label{nullaction}
S[\xi] = \int d \lambda \left [ \frac{1}{2e'} \left(\delta_{ab}\frac{d \xi^a}{d \lambda} \frac{d \xi^b}{d\lambda} -R_{uaub} \xi^a \xi^b\right ) \right ]\,.
\end{align}
The equation of motion of $\xi^a$ now gives
\begin{align}\label{eq_first term}
\frac{d^2\xi^a}{d\lambda^2}=-\tensor{R}{^a_{ubu}}\xi^b\,.
\end{align} 
This is precisely the geodesic deviation equation, consistent with \cite{Witten:2019qhl}, which we have here derived directly from the action as an Euler-Lagrange equation. Note that $a = 1, 2$ in accordance with the fact that the geodesic deviation for null vectors is a spacelike vector that lives in a codimension-2 hypersurface.

In order to quantize the metric, it will be helpful to express everything in Cartesian coordinates and transverse-traceless (TT) gauge. To leading order, the Einstein-Hilbert action in TT gauge is
\begin{align}\label{eq_h action}
    S_{\text{E-H}}\left[h_{\mu \nu}\right]=-\frac{1}{4\kappa^2} \int d^4 x \partial_\mu h_{i j} \partial^\mu h^{i j}\,,
\end{align}
where $\kappa^2=16 \pi G_N$. We can decompose $h_{ij}$ into the Fourier modes:
\begin{align}
\label{h_ij}
    h_{i j}\left(x^i, t\right)=\frac{\kappa}{\sqrt{V}} \sum_{\mathbf{k}, s} h_{\mathbf{k}}^s(t) e^{i \mathbf{k} \cdot \mathbf{x}} e_{i j}^s(\mathbf{k})\,,
\end{align}
where we are working in a finite box $V=L_xL_yL_z$ so that the discrete modes are $k_i=\frac{2\pi n_i}{L_i}$ for $i=1,2,3$. Here, we defined the polarization tensor $e_{ij}^s(\mathbf{k})$ with its normalization $e_{i j}^{* s}(\mathbf{k}) e_{i j}^{s'}(\mathbf{k})=\delta^{s s'}$, and we used the linear polarization for index $s=\times, +$.

We then would like to express our null geodesic deviation equation in Cartesian coordinates. By using the coordinate transformation $u=t+z, v = t-z$, the left hand side in \eqref{eq_first term} becomes
\begin{align}
    \frac{d^2 \xi^a}{du^2} = \frac{d^2 \xi^a}{dt^2}\,,
\end{align}
where we used $\frac{d \xi^a}{dt}=\frac{d \xi^a}{dz}$ since $\xi(u)=\xi(t+z)$.
Due to the coordinate transformation, the components of the Riemann tensor in \eqref{eq_first term} becomes
\begin{align}
    \tensor{R}{^a_{ubu}} = \frac{\partial x^\mu}{\partial u}\frac{\partial x^\nu}{\partial u}\tensor{R}{^a_{\mu b \nu}}\,,
\end{align}
One can compute the components of the Riemann tensor in the linearized gravity. Since $h_{\mu\nu}=h_{\mu\nu}(u)$ along the particle world-line, we obtain
\begin{align}
    \tensor{R}{^{a}_{t b t}}=\tensor{R}{^{a}_{t b z}} =\tensor{R}{^{a}_{z b t}} = \tensor{R}{^{a}_{z b z}} = -\frac{1}{2}\tensor{\ddot{h}}{^a_b}\,,
\end{align}
where $\dot{}$ denotes a derivative with respect to the coordinate time $t$. Combining these, the geodesic deviation equation \eqref{eq_first term} becomes 
\begin{align}
\label{eom cl}
    \ddot{\xi}^a=\frac{1}{2} \tensor{\ddot{h}}{^a_b}\,\xi^b\,,
\end{align}
where $\tensor{h}{^a_b}$ is in the transverse-traceless (TT) gauge. 

Similarly, from \eqref{eq_h action} and \eqref{nullaction}, one can find the equation of motion of the Fourier components of the metric perturbation:
\begin{align}\label{eq_h eom}
    \ddot{h}^s(\mathbf{k}, t)+k^2 h^s(\mathbf{k}, t)=\frac{\kappa}{ 2e' \sqrt{V}} e_{a b}^{*s}(\mathbf{k}) \frac{d^2}{d t^2}\left(\xi^a(t) \xi^b(t)\right)\,,
\end{align}
where $e'$ is the einbein, and $e_{ab}^{*s}$ is a complex conjugate of the polarization tensor. The right-hand side is a source term which gives the backreaction of the fluctuating separation of the null particles on the metric perturbation. In turn, its effect on the geodesic deviation equation will be to introduce a fifth derivative radiation reaction term \cite{Parikh:2020fhy}. We will ignore this term hereafter, as it is unimportant except for rapidly varying changes in the deviation.

\subsection{Canonical quantization and Langevin equation}

Now we would like to quantize this, by promoting the metric to an operator:
\begin{align}
    \hat{g} _{\mu\nu}(x)=\eta_{\mu\nu}+\hat{h}_{\mu\nu}(x)
\end{align}
Following \cite{Kanno:2020usf}, let us promote $h^s(\mathbf{k},t)$ to a quantum operator so that 
\begin{align}
\label{h int}
    \hat{h}^s(\mathbf{k},t) = \hat{a}_s(\mathbf{k})u_k(t) + \hat{a}^\dag_s(\mathbf{-k})u^*_k(t),
\end{align}
where the creation and annihilation operators follow the usual commutation relations,
\begin{align}
    [\hat{a}_s(\mathbf{k}),\hat{a}^\dag_{s'}(\mathbf{k'})] = \delta_{\mathbf{k}\mathbf{k'}} \delta_{s s'}, \,\,\,\,\,\,\,\,\,\, [\hat{a}_s(\mathbf{k}),\hat{a}_{s'}(\mathbf{k'})] = [\hat{a}^\dag_s(\mathbf{k}),\hat{a}^\dag_{s'}(\mathbf{k'})] = 0\,.
\end{align}
Since $\xi^a$ depends on $h_{ab}$, as shown by equation (\ref{eom cl}), then $\xi^a$ must also be promoted to a quantum operator. Formally, the equations of motion of $\hat{\xi}^a$ are in the Heisenberg picture, while equation (\ref{h int}) is in the interaction picture, but since we are ignoring the radiation reaction terms in equation ({\ref{eom cl}}), the Heisenberg and interaction pictures are equivalent.

Let the quantum state of the gravitational field be $\ket{\Psi}$. We can express the quantum operator $\hat{h}$ as the sum of a classical part and a quantum part (where we omit the identity operator multiplying the classical part):
\begin{align}
\label{h decomposition}
    \hat{h} \equiv h_\text{cl}^\Psi + \delta\hat{h}^\Psi\,,
\end{align}
where we defined the classical part as an expectation value of the metric perturbation with respect to a certain quantum state
\begin{align}
h_\text{cl}^\Psi (\mathbf{k},t) \equiv \langle \Psi | \hat{h}(\mathbf{k},t)| \Psi \rangle\,.
\end{align}
With this decomposition, we have taken the quantum fluctuations to be zero on average: $ \langle\delta \hat{h}^\Psi\rangle = 0$. Similarly, $\hat{\xi}^a(t)$ can be decomposed as 
\begin{align}
    \hat{\xi}^a(t) = \xi^a_0(t) + \delta \hat{\xi}^a(t)\,.
\end{align}
We will assume that the classical part $\xi^a_0(t)$ is much larger than the quantum fluctuations $\delta \hat{\xi}^a(t)$. Inserting these in the equation of motion (\ref{eom cl}) and using \eqref{eq_h eom}, we obtain  
\begin{align}
\label{full eom quantum}
    \quad &\ddot{\xi}_0^a (t)+ \delta \ddot{\hat{\xi}}^a (t) = \delta^{a b}\left[ \frac{1}{2} \ddot{h}^\text{cl}_{bc} (t) + 2 \hat{\mathcal{N}}_{bc} (t) \right] \left(\xi^c_0(t) + \delta \hat{\xi}^c(t)\right), 
\end{align}
where 
\begin{align}
    h_{bc}^\text{cl}\left( t\right)& =\frac{ \kappa}{\sqrt{V}} \sum_{\mathbf{k}, s}   e_{b c}^s(\mathbf{k}) h_\text{cl}^s(\mathbf{k},t)\,,\\
\label{noise tensor}
    \hat{\mathcal{N}}_{bc}(t) &=  -\frac{\kappa}{4\sqrt{V}} \sum_s \sum_{\mathbf{k} \leq \Omega_{\mathrm{m}}} k^2  e_{b c}^s(\mathbf{k}) \delta \hat{h}_s^\Psi(\mathbf{k}, t)\,,
\end{align}
for $k=|\mathbf{k}|$. Since we are interested in the quantum-gravitational contribution to the geodesic deviation, we will assume for simplicity that the background does not depend on time, so there is no classical contribution. Then (\ref{full eom quantum}) becomes
\begin{align}
\label{eom quantum}
    \delta \ddot{\hat{\xi}}^a (t) = 2 \delta^{ab} \hat{\mathcal{N}}_{bc} (t) \left(\xi^c_0 (t) + \delta \hat{\xi}^c(t)\right)\,.
\end{align}
Notice that $\ddot{\xi}^a_0$ is zero since we imposed that the classical part of the metric is constant in time. Because of this, any change in the motion of $\hat{\xi}^a(t)$ is only due to the quantum fluctuations described by the above equation. 

Since $\delta \hat{h}$ are the quantum fluctuations of the gravitational field, we can interpret $\hat{\mathcal{N}}_{ab}$ heuristically as the noise in the geodesic separation due to gravitons. In addition, since the quantum fluctuations in the gravitational field have an expectation value of zero, it follows that
\begin{align}\label{eq_noise tensor avg}
    \left\langle \Psi \left| \hat{\mathcal{N}}_{ab}(t)\right | \Psi \right\rangle=0\,,
\end{align}
where, in the operator formalism we are using here, $\langle\cdot\rangle$ now indicates the expectation value of a quantum mechanical operator, not the statistical average of a stochastic function \cite{Bak:2022oyn}. 
The auto-correlation function, or noise kernel, is given by the two-point correlation function of the noise tensors \cite{Kanno:2020usf}
\begin{align}\label{eq_noise kernel}
     K_{abcd}^{\Psi}\left(t, t^{\prime}\right) \equiv \left\langle \Psi \left| \left\{\hat{\mathcal{N}}_{ab}(t) ,\hat{\mathcal{N}}_{cd}\left(t^{\prime}\right)\right\}\right | \Psi \right\rangle = \frac{\kappa^2}{80 \pi^2}\left( \delta_{ac}\delta_{bd}+\delta_{ad}\delta_{bc}-\frac{2}{3}\delta_{ab}\delta_{cd}\right) F^{\Psi}(t,t')\,,
\end{align}
where we used the anticommutator\footnote{$\{\hat{A},\hat{B}\}=\frac{1}{2}(\hat{A}\hat{B}+\hat{A}\hat{B})$} since two noise operators do not commute. Since $F^{\Psi}(t,t')$ is calculated by acting the creation and annihilation operators contained in $\hat{h}$ on the state $\ket{\Psi}$, its form depends is state-dependent (as indicated by the superscript $\Psi$). In addition, its time dependence is $(t-t')$ if we assume time-translation symmetry. The pre-factor is a result of the angular integration of $\mathbf{k}$ with isotropy, and is the same for all rotationally symmetric states.

In general, $F(t,t')$ is a divergent function which needs to be regulated. One way to regularize the function is to introduce an ultraviolet cut-off frequency $\Lambda$. There is a reason for imposing this ultraviolet cut-off. We previously expanded the metric using Fermi normal coordinates. This expansion is convergent only if the wavelength of the gravitational field $h_{\mu \nu}$ is larger than the characteristic length scale of the system $\xi$.

The cut-off is determined by the length scale of the deviation between two geodesics. This relation comes from the assumption that we take when we use the Fermi normal coordinates. One can understand this relation in the following way. The Fermi normal coordinates in \eqref{eq_FNC1} imply the condition $|\ddot{h} \,\xi_0^2| \ll \mathcal{O}(1)$ where $\xi_0$ is the length of the initial geodesic deviation vector. If we only keep track of the $\omega$ dependence in \eqref{h_ij}, we have
\begin{align}
    &|h(0)| \sim \left|\int^\Lambda_0 d\omega \omega^2 A(\omega)\right|\,,\\
    &|\ddot{h}(0)| \sim \left|\int^\Lambda_0 d\omega \omega^4 A(\omega)\right|\,,
\end{align}
where $A(\omega)$ is an amplitude for each modes $\omega$. For the Poincar\'e-invariant (``Minkowski") vacuum state, $|0 \rangle$, we have $A(\omega)\sim \omega^{-1/2}$ so that $|\ddot{h}(0)|\sim |h(0)\Lambda^2|$. Since we have $h(0) \ll 1$ for the linearized perturbation, the assumptions in the Fermi normal coordinate $|\ddot{h} \,\xi_0^2| \ll \mathcal{O}(1)$ implies $\Lambda \sim \xi_0^{-1}$. In this sense, the UV cut-off is determined by the length scale of the geodesic deviation.

In the Minkowski vacuum, $F(t,t')$ is obtained as \cite{Kanno:2020usf}
\begin{align}\label{eq_F tt}
    F^{|0 \rangle}[\Lambda(t-t')]= \int_0^\Lambda \, dk k^5 \cos[k (t-t')]\,,
\end{align}
where we used the fact that the mode function in equation (\ref{h int}) in Mikowski vacuum is given by
\begin{align}
\label{wave modes}
    u_k(t) = \frac{1}{\sqrt{2k}}e^{ikt}\,.
\end{align}

The auto-correlation function in the Minkowski vacuum can be calculated by using the definition in \eqref{eq_noise kernel} with \eqref{eq_F tt}. The auto-correlation function, which is essentially a two-point function of the gravitational field operators, has a UV divergence. To see this, let us consider how the auto-correlation function changes if we don't impose the ultraviolet cut-off, but using point-splitting regularization. Suppose that we had considered modes of all frequencies. Then, the factor of $e^{i \mathbf{k} \cdot \mathbf{x}}$ from equation (\ref{h_ij}) would be present in the definition of the noise tensor (\ref{noise tensor}) (which would now also depend on space). The auto-correlation function would have the same definition as before (\ref{eq_noise kernel}), except that it would now depend on space in addition to time. Working in the Minkowski vacuum case, and absorbing the extra factor of $e^{i \mathbf{k} \cdot \mathbf{x}}$ into the wave modes (\ref{wave modes}), it is straightforward to show that 
\begin{equation}
    K_{abcd}^{|0\rangle}\left(x^\mu, x^{\prime \mu}\right) \equiv \left\langle 0 \left| \left\{\hat{\mathcal{N}}_{ab}(x^\mu) ,\hat{\mathcal{N}}_{cd}\left(x^{\prime \mu}\right)\right\}\right | 0 \right\rangle \sim \int_0^\infty dk \, k^5 \cos[\eta_{\mu \nu} k^\nu (x^\mu - x^{\prime \mu})]\,,
\end{equation}
where we have omitted a few constants, as well as the polarization tensors, and the integral over the angles. The integral involving the high-frequency modes can be written as
\begin{equation}
    \int_\Lambda^\infty d k \, k^5 \left(\cos[k(t-t^\prime)] \cos[\Vec{k} \cdot(\Vec{x}-\Vec{x}^\prime)]+\sin[k(t-t^\prime)] \sin[\Vec{k}\cdot(\Vec{x}-\Vec{x}^\prime)]\right)\,.
\end{equation}
If the distance between two points $|\Vec{x}-\Vec{x}^\prime|$ is not significantly smaller than the cut-off wavelength $\frac{2\pi}{\Lambda}$, we infer that the integrand for equal time must oscillate at very high frequencies for $k\gg\Lambda$, meaning that they give no net contribution to the above integral. This idea is very similar to how path integrals can be approximated using saddle points. Thus, to a good approximation, rather than integrating over all frequencies we may impose an ultraviolet cut-off. As noted above, the natural value for the cut-off is $\Lambda \sim \xi_0^{-1}$.

\subsection{Quantum fluctuations in the geodesic deviation equation}
We will solve the equations of motion (\ref{eom quantum}) perturbatively. Let 
\begin{align}
    \delta \hat{\xi}^a(t) = \hat{\xi}_1^a(t) + \hat{\xi}_2^a(t) + \dots,
\end{align}
where the subscript indicates the power of the quantum fluctuations $\delta \hat{h}_s^{\Psi}$, equivalently the power of $\hat{\mathcal{N}}_{ab}$. For example, the classical part $\xi^a_0$ is the zeroth order term, while $\hat{\xi}_1^a(t)$ is the first order quantum correction to $\hat{\xi}^a(t)$. This approach leads to ambiguities in the higher-order terms due to the fact that our operators do not commute. We did not encounter this issue when we performed a similar calculation working in the influence functional formalism \cite{Bak:2022oyn}, since in that case our equations of motion contained only c-numbers that commuted with each other. Thus, for consistency with our previous results, we will introduce an anticommutator in the equations of motion (\ref{eom quantum}), which is equivalent to assuming Weyl ordering. This gives
\begin{align}
    \delta \ddot{\hat{\xi}}^a (t) = 2 \delta^{ab} \left\{\hat{\mathcal{N}}_{bc} (t), \hat{\xi}^c(t)\right\}\, ,
\end{align}
where $\hat{\xi}^a(t)=\xi^a_0(t)+\delta\hat{\xi}^a(t)$. Up to second order, this gives us
\begin{align}
    \ddot{\xi}^a_0(t)&=0\,,\\
    \ddot{\hat{\xi}}^a_1 (t) &= 2  \delta^{ab} \left\{\hat{\mathcal{N}}_{bc} (t),  \xi^c_0(t) \right\} \,,\\
    \ddot{\hat{\xi}}^a_2 (t) &=  2 \delta^{ab}  \left\{ \hat{\mathcal{N}}_{bc} (t), \hat{\xi}^c_1 (t) \right\}\,.
\end{align}
To avoid cumbersome notation, we will avoid writing the anticommutators explicitly in the following calculations; however, whenever we have a product of two operators, it should be understood as the anticommutator of the two operators. 

Putting everything together, we have
\begin{align}
    \begin{aligned}
\xi^a(t)=\xi_0^a(t) & +2 \delta^{ab} \int_0^t d \tau_1 \int_0^{\tau_1} d \tau_2 \hat{\mathcal{N}}_{bc}\left(\tau_2\right) \xi_0^c (\tau_2)\\
& +4 \delta^{ab} \delta^{cd} \int_0^t d \tau_1 \int_0^{\tau_1} d \tau_2 \int_0^{\tau_2} d \tau_3 \int_0^{\tau_3} d \tau_4 \hat{\mathcal{N}}_{bc}\left(\tau_2\right) \hat{\mathcal{N}}_{de}\left(\tau_4\right) \xi_0^e(\tau_4)+\cdots\,.
\end{aligned}
\end{align}
One point function of the first-order term vanishes since the average of the noise operator $\hat{\mathcal{N}}_{ab}$ is zero \eqref{eq_noise tensor avg}.
\begin{align}
    \langle\Psi| \hat{\xi}_1^a(t)|\Psi\rangle=0 \,.
\end{align}
However, the one-point function of the second-order term includes the auto-correlation function, so it does not vanish.
\begin{align}
    \langle\Psi| \hat{\xi}_2^a(t)|\Psi\rangle=4 \delta^{ab} \delta^{cd} \int_0^t d \tau_1 \int_0^{\tau_1} d \tau_2 \int_0^{\tau_2} d \tau_3 \int_0^{\tau_3} d \tau_4 K_{bcde}^{\Psi}\left(\tau_2, \tau_4\right) \xi_0^e(\tau_4)
\end{align}
There is also a non-zero two-point function of the first-order terms.
\begin{align}\label{eq_first two point}
    \langle \Psi|\hat{\xi}_1^a(t) \,\hat{\xi}_1^b(t^{\prime})|\Psi\rangle=4 \delta^{ac} \delta^{bd} \int_0^t d \tau_1 \int_0^{\tau_1} d \tau_2 \int_0^{t^{\prime}} d \tau_3 \int_0^{\tau_3} d \tau_4 K_{cedf}^{\Psi}\left(\tau_2, \tau_4\right) \xi_0^e\left(\tau_2\right) \xi_0^f(\tau_4)
\end{align}

\begin{figure}
    \centering
    \includegraphics[width=0.52\textwidth]{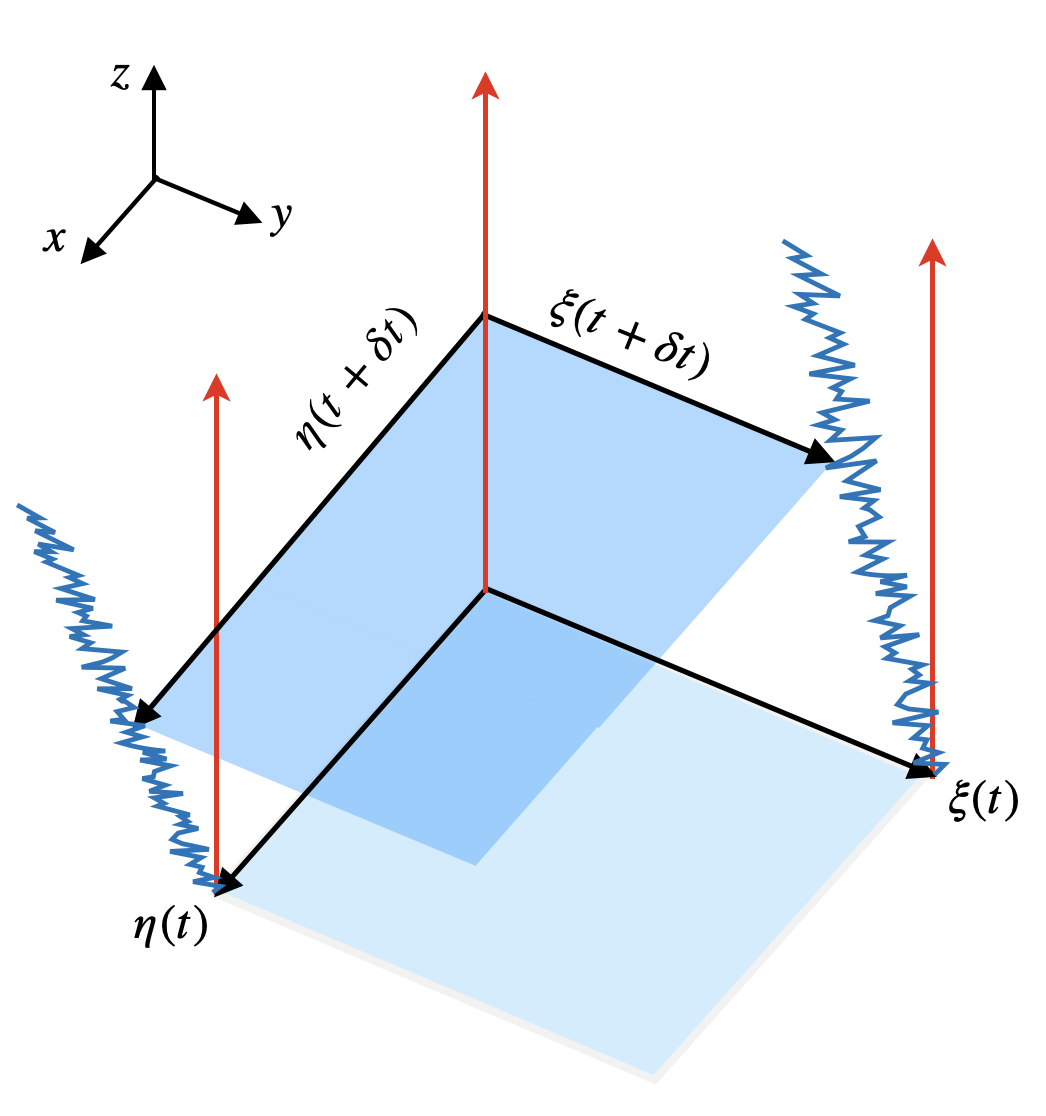}
    \caption{The diagram shows the evolution of a null congruence in three-dimensional space. The red arrows indicate the classical trajectories of null particles moving in the $z$-direction. The black arrows indicate the geodesic deviation vectors influenced by the quantum fluctuations of spacetime. Quantum fluctuations cause the geodesic separation to have a stochastic motion, with one realization schematically portrayed by the blue lines. The time evolution of the area of the congruence is represented by the blue sheets. Because of correlations in the fluctuations in different directions, the area change can have a definite sign.}
    \label{fig-congruence}
\end{figure}

\section{Null Congruences in a Quantized Gravitational Field}\label{sec-quantum correction}

Having determined how quantum-gravitational fluctuations affect the geodesic deviation equation, we turn now to the problem of interest, which is to calculate the quantum-gravitational evolution of the area of a future-directed null geodesic congruence. Consider a pencil of light-like geodesics with rectangular transverse area; this could be the beam of a rectangular headlight. This area can be expressed in terms of a product of geodesic deviations as follows. Choose the geodesic at one of the corners of the rectangle as a reference geodesic. Consider two other geodesics at the two adjacent corners of the rectangle. Then, the length and width of the rectangle correspond to the geodesic deviations of these geodesics from the reference geodesic; see Figure~\ref{fig-congruence}.

The area of the rectangle is therefore just the product of the geodesic distances, so we can define the area operator:
\begin{align}
   \mathcal{\hat{A}}(t)=\hat{\xi}(t) \hat{\eta}(t)\,,
\end{align}
where again the product is to be understood as an anticommutator: $\mathcal{\hat{A}}(t)= \{ \hat{\xi}(t) , \hat{\eta}(t) \}$.
For simplicity, choose the basis vectors in the codimension$-2$ subspace to lie along $\xi$ and $\eta$, so that they have only nonzero spatial component. If the geodesics have no shear or initial expansion, then the classical Raychaudhuri equation indicates that this transverse area will stay constant: $\mathcal{A}_c = \xi_0 \eta_0$. However, as we have seen, quantum-gravitational fluctuations make the geodesic deviation operators  evolve via the quantum geodesic deviation, (\ref{eom quantum}). This can be solved perturbatively. 

Expanding the geodesic deviation vectors in order of the noise terms as
\begin{align}
    \begin{aligned}
& \hat{\xi}(t)=\xi_0+\hat{\xi}_1(t)+\hat{\xi}_2(t)+\cdots \,,\\
& \hat{\eta}(t)=\eta_0+\hat{\eta}_1(t)+\hat{\eta}_2(t)+\cdots\,.
\end{aligned}
\end{align}
Then, taking the expectation value of the area operator, we find that
\begin{align}
\label{avgA}
    \begin{aligned}
\langle\Psi| \mathcal{\hat{A}}(t)|\Psi\rangle & =\xi_0 \eta_0 +\langle\Psi| \hat{\xi}_1\hat{\eta}_1|\Psi \rangle+\eta_0 \langle\Psi|\hat{\xi}_2|\Psi\rangle+\xi_0\langle\Psi|\hat{\eta}_2|\Psi\rangle+\cdots \\
& \equiv \mathcal{A}_c+\mathcal{A}_q(t)\,.
\end{aligned}
\end{align}
Even though the noise averages to zero, (\ref{eq_noise tensor avg}), we see that the expectation value of the quantum-gravitational contribution to the area does not vanish because it picks up the two-point function of geodesic deviations, as well as the one-point function of the second-order deviations.

\subsection{Quantum-gravitational Raychaudhuri equation}
In order to determine the quantum-corrected time evolution of the area, we see from \eqref{avgA} that we need to first compute the two-point function and one-point function. These depend on $|\Psi \rangle$, the quantum state of the gravitational field. To be concrete, we will hereafter take the state to be the Poincar\'e-invariant vacuum state, $|0 \rangle$. 

Then, using \eqref{eq_noise kernel}, \eqref{eq_first two point} and \eqref{eq_F tt}, we find that
\begin{align}
    \langle 0 | \hat{\xi}_1(t)\, \hat{\eta}_1(t) | 0 \rangle=-\frac{\kappa^2\xi_0 \eta_0 }{30 \pi^2} \int_0^t d \tau_1 \int_0^{\tau_1} d \tau_2 \int_0^{t} d \tau_3 \int_0^{\tau_3} d \tau_4 F^{|0 \rangle}\left(\tau_2, \tau_4\right) \simeq-\frac{\kappa^2 \xi_0 \eta_0}{120 \pi^2}\Lambda^4 t^2\,.
\end{align}
The minus sign of the two-point function implies that there is a negative correlation between two different geodesic deviations. If one of the lengths is shrinking, then the other one should be expanding. 
Similarly, we obtain the one-point function of the second-order terms:
\begin{align}
    \begin{aligned}
& \langle 0 | \hat{\xi}_2(t)|0 \rangle= \frac{\kappa^2 \xi_0}{6 \pi^2} \int_0^t d \tau_1 \int_0^{\tau_1} d \tau_2 \int_0^{\tau_2} d \tau_3 \int_0^{\tau_3} d \tau_4 F^{|0 \rangle}\left(\tau_2, \tau_4\right) \simeq-\frac{\kappa^2 \xi_0}{48 \pi^2 }\Lambda^4 t^2 \\
& \langle 0 | \hat{\eta}_2(t)| 0 \rangle= \frac{\kappa^2\eta_0}{6\pi^2} \int_0^t d \tau_1 \int_0^{\tau_1} d \tau_2 \int_0^{\tau_2} d \tau_3 \int_0^{\tau_3} d \tau_4 F^{|0 \rangle}\left(\tau_2, \tau_4\right) \simeq-\frac{\kappa^2 \eta_0}{48 \pi^2 }\Lambda^4 t^2
\end{aligned}
\end{align}
Unlike the two-point function, the angular pre-factor in \eqref{eq_noise kernel} results here in an overall plus sign. 

Thus, by \eqref{avgA}, the quantum correction to the area is
\begin{align}
    \mathcal{A}^{|0\rangle}_q(t)\simeq-\frac{\kappa^2 }{20 \pi^2}\mathcal{A}_c\,\Lambda^4 t^2\,.
\end{align}
We define the quantum expansion of the null congruence via 
\begin{align}\label{eq_quantum expansion}
    \Theta(t) \equiv \frac{1}{\langle\hat{\mathcal{A}}(t)\rangle}\frac{d\langle{\hat{\mathcal{A}}}(t)\rangle}{dt}\,.
\end{align}
By taking the time derivative and keeping track of the leading order corrections in $A_q/A_c$, the expansion is written as
\begin{align}
    \Theta(t)=\frac{\dot{\mathcal{A}}_c(t)}{\mathcal{A}_c(t)}+\frac{d}{d t}\left(\frac{\mathcal{A}_q(t)}{\mathcal{A}_c(t)}\right) \equiv \theta_c(t)+\theta_q(t) \,,
\end{align}
where, for the simple case of a static classical area, the quantum correction to the expansion is
\begin{align}
    \theta^{|0 \rangle}_q(t)\simeq-\frac{\kappa^2}{10 \pi^2}\Lambda^4 t\,.
\end{align}
Then, differentiating again, we find
\begin{align}\label{eq_Quantum Raychaudhuri}
    \dot{\Theta}^{|0 \rangle}(t)=-\frac{1}{2} \theta_c^2-\sigma^2-R_{\mu \nu} k^\mu k^\nu-\frac{\kappa^2}{10 \pi^2}\Lambda^4\,.
\end{align}
This is the quantum-gravity corrected Raychaudhuri equation when the gravitational field is in the Poincar\'e-invariant vacuum state. Here we have allowed the classical area to be time-dependent, so that $\theta_c$ obeys the classical Raychaudhuri equation, and we have set the vorticity term, $\omega^2$, to zero by hypersurface-orthogonality. We see that the effect of quantum fluctuations of spacetime is to introduce an additional term on the right-hand side of Raychaudhuri's equation. The term depends on $\Lambda$, the frequency cut-off scale which, as noted, is inversely proportional to the characteristic size of the congruence. Notably, the new term has -- at least for the vacuum state -- the same negative sign as the usual classical terms. It would be very interesting to determine the sign of this term for other classes of states,  to see whether a general statement can be made.

\section{A Quantum-Gravitational Focusing Theorem} \label{sec-focus}

\paragraph{Classical Focusing Theorem} Consider the classical Raychaudhuri equation for a null congruence: 
\begin{align}
    \dot{\theta}_c = -\frac{\theta_c^2}{2}-\sigma^2 -R_{\mu\nu}k^\mu k^\nu
\end{align}
where again we have discarded the vorticity term 
by hypersurface-orthogonality. 
By Einstein's equations, we have $R_{\mu\nu}k^\mu k^\nu = \frac{\kappa^2}{2} T_{\mu\nu}k^\mu k^\nu$. 
If we now impose the null energy condition $T_{\mu\nu}k^\mu k^\nu\geq 0$, the null Raychaudhuri equation implies that the classical expansion of the congruence always decreases, i.e.,
\begin{align}
    \dot{\theta}\leq 0\,.
\end{align}
This is known as the classical focusing theorem: it implies that the expansion must decrease during the congruence’s evolution. Thus, an initially diverging congruence will diverge less rapidly in the future, while an initially converging congruence will converge more rapidly in the future. This theorem is quite generic in general relativity for which, of course, there is a definite classical geometry. However, to the best of our knowledge, it has not yet been investigated how the classical focusing theorem is modified in situations where quantum fluctuations of the metric are taken into account.

\paragraph{A Quantum Focusing Theorem} In this work, we have derived a quantum-gravitational modification of the Raychaudhuri equation for a pencil of null geodesics \eqref{eq_Quantum Raychaudhuri}, simply
\begin{align}\label{eq_Quantum Raychaudhuri 2}
    \dot{\theta}_c +\dot{\theta}_q = -\frac{\theta_c^2}{2}-\sigma^2 -R_{\mu\nu}k^\mu k^\nu-\frac{4 \hbar G_N }{\mathcal{A}_0}\left(\frac{32\pi^3}{5 \mathcal{A}_0}\right)\,,
\end{align}
where we used $\mathcal{A}_0 \sim\left(\frac{2\pi}{\Lambda}\right)^2$ and have explicitly restored $\hbar$. The first three terms on the right-hand side of the equation \eqref{eq_Quantum Raychaudhuri 2}  originate from the classical contribution, while the last term involves a quantum-gravitational contribution. It is important to note that this quantum correction comes from the quantum nature of spacetime itself, rather than from quantum matter fields. 

Interestingly, the strength of the quantum-gravitational fluctuation depends on the characteristic length scale of the geodesic congruence: the quantum fluctuation of spacetime increases when the scale of a pencil of geodesics decreases. This scale dependence seems counter-intuitive at first glance. For instance, it means that if we divide a congruence of geodesics into sub-congruences, then those smaller sub-congruences fluctuate more than the whole system. In particular, the sum of the quantum-corrected areas of the sub-congruences exceeds the quantum-corrected area of the original congruence. This fractal scaling resembles the length of a coastline, which increases with resolution; it is also reminiscent of previous work proposing that quantum gravity induces a fractal structure to spacetime \cite{Knizhnik:1988ak}.

If we take into account only classical matter, we can use the Einstein equation $R_{\mu\nu}k^\mu k^\nu=\frac{\kappa^2}{2} T_{\mu\nu}k^\mu k^\nu$ and the null energy condition $T_{\mu\nu}k^\mu k^\nu\geq 0$. Remarkably, the quantum-gravitational correction has a negative sign, so that the right-hand side of \eqref{eq_Quantum Raychaudhuri 2} is always negative. Therefore, the quantum-gravitational null Raychaudhuri equation also implies a focusing theorem:
\begin{align}
\label{quantum focusing thm}
    \dot{\Theta}^{|0 \rangle} \leq 0\,,
\end{align}
where $\Theta=\theta_c+\theta_q$ with the definition \eqref{eq_quantum expansion}. It appears that, at least for the Minkowski vacuum state,  perturbative quantum corrections do not spoil the classical focusing theorem. This also suggests the formation of a caustic in the future, as in the case of timelike geodesics \cite{Bak:2022oyn}. It would be interesting to study how this affects the black hole singularity theorems.

If we take into account the effect of quantized matter fields and examine the contribution from the renormalized energy-momentum tensor within a semi-classical framework, we can obtain a deviation from the classical null energy condition \cite{Birrell:1982ix}. This results in a potential alteration of the sign in the Raychaudhuri equation. What we have found is that, perhaps surprisingly, the effect of linearized perturbative quantum gravity is qualitatively different. The quantum fluctuation of the metric simply contributes a term that aligns with the sign of the classical contributions.

There is also a quantum focusing conjecture motivated by the holographic principle and generalized second law \cite{Bousso:2015mna}, which is related to \cite{Bousso:2015wca,Bousso:2022tdb}. In this conjecture, a quantum expansion is introduced, incorporating a contribution from quantum matter fields in the generalized entropy associated with a holographic screen. The aim is to extend the focusing theorem to scenarios where quantum matter might violate the null energy condition and the validity of the generalized second law. A quantum null energy condition \cite{Bousso:2015wca} was proposed as a sufficient condition for the quantum focusing conjecture (similarly, the focusing theorem can also be motivated on thermodynamic grounds \cite{Parikh:2015ret}). Our conjecture differs from their framework in several aspects. First, in our approach, the quantum correction arises from perturbative quantum gravity, and we consider the expansion of a pencil of the congruence rather than a holographic screen, representing the boundary of certain subregions. Second, we assume that the matter sector continues to obey the usual null energy condition. However, despite these differences, the ultimate outcome in both situations appears to be linked by the non-violation of classical focusing of null geodesics in the presence of quantum effects. Therefore, it might be an interesting future work to investigate a setup that has an overlap between the two works. \\

\section{Summary} \label{summary}

In this paper, we have considered the effect of spacetime fluctuations on the deviation of null geodesics in the regime of linearized perturbative quantum gravity, and used that to obtain a quantum-gravitational null Raychaudhuri equation. Our equation has an additional term on the right which originates in stochastic fluctuations of the geodesic separation induced by the quantum fluctuations of the metric. The details of the new term depend on the quantum state of the gravitational field. Remarkably, when the gravitational field is in the usual vacuum state, the sign of the correction is the same as that of the classical terms and therefore does not significantly alter the conclusion of the classical focusing theorem. It would be very interesting to be able to prove that this true in general for all quantum states of the gravitational field. But if that were to be true, then presumably the singularity theorems -- which rely crucially on the Raychaudhuri equation -- would also continue to hold, and linearized quantum gravity at least would then be unable to save general relativity from the menace of singularities. \\

\noindent
{\bf Acknowledgments}\\
We thank Samarth Chawla, Jude Pereira, and George Zahariade for helpful discussions. MP is supported in part by Heising-Simons Foundation grant 2021-2818, Department of Energy grant DE-SC0019470 and Government of India DST VAJRA Faculty Scheme
VJR/2017/000117. The research of S.S. is supported by the Department of Science and Technology, Government of India under the SERB
CRG Grant CRG/2020/004562.


\bibliographystyle{JHEP}

\end{document}